\documentclass[ pra,letter,  paper,superscriptaddress]{revtex4-2}
\usepackage{graphicx,amsmath,amssymb,amsfonts,latexsym,xcolor,dcolumn,bm}
\usepackage{appendix}
\usepackage[colorlinks]{hyperref}
\usepackage{color}

\begin{document}

\global\long\def\eqn#1{\begin{align}#1\end{align}}
\global\long\def\vec#1{\overrightarrow{#1}}
\global\long\def\ket#1{\left|#1\right\rangle }
\global\long\def\bra#1{\left\langle #1\right|}
\global\long\def\bkt#1{\left(#1\right)}
\global\long\def\sbkt#1{\left[#1\right]}
\global\long\def\cbkt#1{\left\{#1\right\}}
\global\long\def\abs#1{\left\vert#1\right\vert}
\global\long\def\cev#1{\overleftarrow{#1}}
\global\long\def\der#1#2{\frac{{d}#1}{{d}#2}}
\global\long\def\pard#1#2{\frac{{\partial}#1}{{\partial}#2}}
\global\long\def\re{\mathrm{Re}}
\global\long\def\im{\mathrm{Im}}
\global\long\def\dd{\mathrm{d}}
\global\long\def\ddd{\mathcal{D}}
\global\long\def\avg#1{\left\langle #1 \right\rangle}
\global\long\def\mr#1{\mathrm{#1}}
\global\long\def\mb#1{{\mathbf #1}}
\global\long\def\mc#1{\mathcal{#1}}
\global\long\def\tr{\mathrm{Tr}}
\global\long\def\dbar#1{\stackrel{\leftrightarrow}{\mathbf{#1}}}

\global\long\def\nth{$n^{\mathrm{th}}$\,}
\global\long\def\mth{$m^{\mathrm{th}}$\,}
\global\long\def\non{\nonumber}

\newcommand{\bU}{{\bf{U}}}
\newcommand{\bV}{{\bf{V}}}
\newcommand{\bW}{{\bf{W}}}
\newcommand{\bd}{{\bf d}}
\newcommand{\hr}{\hat{\br}}
\newcommand{\bM}{\bf{M}}
\newcommand{\bv}{{\bf v}}
\newcommand{\hbp}{\hat{\bp}}
\newcommand{\hq}{\hat{q}}
\newcommand{\hp}{\hat{p}}
\newcommand{\ha}{\hat{a}}
\newcommand{\had}{{a}^{\dag}}
\newcommand{\ad}{a^{\dag}}
\newcommand{\hsig}{{\hat{\sigma}}}
\newcommand{\nt}{\tilde{n}}
\newcommand{\itf}{\sl}
\newcommand{\eps}{\epsilon}
\newcommand{\bsig}{\pmb{$\sigma$}}
\newcommand{\beps}{\pmb{$\eps$}}
\newcommand{\bmu}{\pmb{$ u$}}
\newcommand{\balpha}{\pmb{$\alpha$}}
\newcommand{\bbeta}{\pmb{$\beta$}}
\newcommand{\bgamma}{\pmb{$\gamma$}}
\newcommand{\bu}{{\bf u}}
\newcommand{\bpi}{\pmb{$\pi$}}
\newcommand{\bSig}{\pmb{$\Sigma$}}
\newcommand{\be}{\begin{equation}}
\newcommand{\ee}{\end{equation}}
\newcommand{\bea}{\begin{eqnarray}}
\newcommand{\eea}{\end{eqnarray}}
\newcommand{\sss}{_{{\bf k}\lambda}}
\newcommand{\ssss}{_{{\bf k}\lambda,s}}
\newcommand{\dip}{\langle\sigma(t)\rangle}
\newcommand{\dipp}{\langle\sigma^{\dag}(t)\rangle}
\newcommand{\sig}{{{\sigma}}}
\newcommand{\sigd}{{\sigma}^{\dag}}
\newcommand{\sigz}{{\sigma_z}}
\newcommand{\ra}{\rangle}
\newcommand{\la}{\langle}
\newcommand{\om}{\omega}
\newcommand{\Om}{\Omega}
\newcommand{\pa}{\partial}
\newcommand{\bR}{{\bf R}}
\newcommand{\bx}{{\bf x}}
\newcommand{\br}{{\bf r}}
\newcommand{\bE}{{\bf E}}
\newcommand{\bH}{{\bf H}}
\newcommand{\bB}{{\bf B}}
\newcommand{\bP}{{\bf P}}
\newcommand{\bD}{{\bf D}}
\newcommand{\bA}{{\bf A}}
\newcommand{\bek}{{\bf e}\rmk}
\newcommand{\rmk}{_{{\bf k}\lambda}}
\newcommand{\rk}{_{{\bf k}_1{\lambda_1}}}
\newcommand{\rkk}{_{{\bf k}_2{\lambda_2}}}
\newcommand{\rkz}{_{{\bf k}_1{\lambda_1}z}}
\newcommand{\rkkz}{_{{\bf k}_2{\lambda_2}z}}
\newcommand{\bsij}{{\bf s}_{ij}}
\newcommand{\bk}{{\bf k}}
\newcommand{\bp}{{\bf p}}
\newcommand{\epso}{{1\over 4\pi\eps_0}}
\newcommand{\bS}{{\bf S}}
\newcommand{\bL}{{\bf L}}
\newcommand{\bJ}{{\bf J}}
\newcommand{\bI}{{\bf I}}
\newcommand{\bF}{{\bf F}}
\newcommand{\bsub}{\begin{subequations}}
\newcommand{\esub}{\end{subequations}}
\newcommand{\baline}{\begin{eqalignno}}
\newcommand{\ealine}{\end{eqalignno}}
\newcommand{\Ep}{{\bf E}^{(+)}}
\newcommand{\Em}{{\bf E}^{(-)}}
\newcommand{\hbx}{{\hat{\bf x}}}
\newcommand{\hby}{{\hat{\bf y}}}
\newcommand{\hbz}{{\hat{\bf z}}}
\newcommand{\bep}{\hat{\bf e}_+}
\newcommand{\bem}{\hat{\bf e}_-}
\newcommand{\orange}[1]{{\color{orange} {#1}}}
\newcommand{\cyan}[1]{{\color{cyan} {#1}}}
\newcommand{\blue}[1]{{\color{blue} {#1}}}
\newcommand{\yellow}[1]{{\color{yellow} {#1}}}
\newcommand{\green}[1]{{\color{green} {#1}}}
\newcommand{\red}[1]{{\color{red} {#1}}}
\newcommand{\pr}{^{\prime}}
\newcommand{\hd}{\hat{\bd}}
\newcommand{\hk}{\hat{\bk}}
\title{Effect of Self-Interaction on Feynman's Interpretation of the Lamb Shift}
\author{Peter W. Milonni}
\affiliation{Department of Physics and Astronomy, University of Rochester, Rochester, New York 14627, USA}
\author{Paul R. Berman}
\affiliation{Physics Department, University of Michigan, Ann Arbor, Michigan 48109-1040, USA}
\author{Kanu Sinha}
\affiliation{College of Optical Sciences and Department of Physics, University of Arizona, Tucson, Arizona 85721, USA}
\begin{abstract}
We derive Bethe's formula for the Lamb shift by extending Feynman's suggestion that the shift could be interpreted as the change, due to the presence of the atom, in  electromagnetic {{field}} energy. This approach is based on measurable quantities such as a refractive index but has a contribution from virtual photon absorption which is effectively eliminated by a high-energy cutoff in the nonrelativistic theory. We show that this unphysical contribution is cancelled when a self-interaction energy is included in Feynman's argument. 
\end{abstract}
\maketitle

\section{Introduction}
Hans Bethe's nonrelativistic calculation of the Lamb shift \cite{bethe} was said by Feynman to be ``the most important discovery in the history of quantum electrodynamics \cite{feyn}.'' The ``Bethe log'' formula has over many years been reviewed and used in numerous books and papers \cite{refs,jent}. The Lamb shift itself continues to be cited as a prime example of the observable effects of vacuum fields and fluctuations in quantum field theory. Feynman suggested that the Lamb shift could be attributed to the change in the electromagnetic field energy resulting from the presence of the atom \cite{feyn2}. He expressed this interpretation in terms of a measurable physical quantity: the forward scattering amplitude, or equivalently the polarizability of the atom or the refractive index of the ``medium" consisting of the one atom. A more detailed derivation of Bethe's formula following this suggestion was carried out by Power \cite{power}. 

Given its fundamental importance, yet another physical interpretation of Bethe's formula might be worthwhile. In this paper we consider, in addition to the change considered by Feynman in the electromagnetic energy caused by an atom, the energy associated with the interaction of the atom with its own field. We assume that the atom is in its ground state in the absence of any externally applied field. 

In the following section we review the Feynman argument and how it involves an energy associated with (unphysical) virtual photon absorption. This unphysical energy is shown to be effectively eliminated by a high-energy cutoff in the nonrelativistic theory. In Section \ref{sec:self} we obtain an expression for the self-interaction of the atom, and in Section \ref{sec:log} it is shown that, when this additional energy is included in Feynman's argument, the energy corresponding to virtual photon absorption is exactly cancelled, independently of the cutoff, and the Bethe log follows after subtraction of the free-electron energy and mass renormalization. We conclude in Section \ref{sec:disc} with some brief remarks.

\section{The Bethe log in Feynman's interpretation}\label{sec:feyn}
We begin with the zero-temperature energy $\sum\sss\frac{1}{2}\hbar\om_k$
of plane-wave electromagnetic modes with wave vectors $\bk$ and polarizations $\lambda$ in a large volume $V$ containing a homogeneous, isotropic dielectric medium with a refractive index $n(\om_k)$ due to a density $N$ of identical ground-state atoms. The wavelengths that can fit in a box of volume $V$ are not changed by the refractive index, but the frequencies are changed from $\om_k$ to $\om_k/n(\om_k)$, consistent with a phase velocity $c/n(\om_k)$ at frequency $\om_k$. The electromagnetic energy is therefore (see also Appendix A)
\be
2\sum_{\bk}\frac{1}{2}\frac{\hbar kc}{n}=\frac{V}{8\pi^3}\int d^3k\frac{\hbar kc}{n},
\label{eqf}
\ee
where we have summed over the two independent polarizations for each $\bk$. Changing variables from $k$ to $\om=kc/n$, this can be written as 
\be
\frac{\hbar V}{8\pi^3c^3}4\pi\int_0^{\Om}d\om\frac{dk}{d\om}(n\om)^2\om=\frac{\hbar V}{2\pi^2c^3}\int_0^{\Om}d\om\om(n\om)^2\frac{d}{d\om}(n\om),
\label{eq1}
\ee
where $\Omega$ is a high-frequency cutoff. The difference between this energy and the energy in the absence of any atoms is
\be
U=\frac{\hbar V}{2\pi^2c^3}\int_0^{\Omega}d\om\big[n^2\om^3\frac{d}{d\om}(n\om)-\om^3\big]=\frac{\hbar V}{2\pi^2c^3}\int_0^{\Omega}d\om(n^3-1)\om^3+\frac{\hbar V}{2\pi^2c^3}\frac{1}{3}\int_0^{\Omega}d\om\om^4\frac{d}{d\om}(n^3).
\ee
By partial integration,
\be
\int_0^{\Omega}d\om\om^4\frac{d}{d\om}(n^3)=\om^4n^3\Big|_0^{\Omega}-4\int_0^{\Omega}d\om\om^3(n^3-1)-4\int_0^{\Omega}d\om\om^3=-4\int_0^{\Omega}d\om\om^3(n^3-1),
\ee
since in the high-frequency limit $n(\Omega)=1$. Therefore,
\be
U=-\frac{\hbar V}{6\pi^2c^3}\int_0^{\Omega}d\om\om^3(n^3-1).
\label{eqq4}
\ee
This compares with the energy
\be
U_P=-\frac{\hbar V}{2\pi^2c^3}\int_0^{\Omega}d\om\om^3(n-1)
\label{eqq5}
\ee
obtained by Power \cite{power}, which did not account for $dn/d\om$ in the sum over modes. The difference between $U$ and $U_P$ is immaterial for $n=1+\delta n$, $\delta n\ll 1$. Then
\be
U\cong U_P\cong -\frac{\hbar V}{2\pi^2c^3}\int_0^{\Omega}d\om\om^3\delta n(\om).
\label{eqexp}
\ee

Before deriving the Lamb shift from (\ref{eqexp}), we note that (\ref{eqexp}) has the appealing feature that it is expressed in terms of {\sl measurable} physical quantities---the refractive index or forward scattering amplitude or  polarizability---even though the Lamb shift is associated with {\sl virtual} processes \cite{feyn3}. However, as discussed below, it has an unphysical feature.

From the Kramers--Heisenberg formula~\cite{refs,KramersHeisenberg},
\be
\delta n(\om)=\frac{4\pi N}{3\hbar}\sum_m\frac{\om_{mg}|\bd_{mg}|^2}{\om_{mg}^2-(\om+i0^+)^2}=2\pi N\alpha_R(\om)
\ee
for a dilute medium of off-resonant ground-state atoms, where $\om_{mg}>0$ is the angular transition frequency between the excited state $m$ and the ground state $g$, $\bd_{mg}$ is the corresponding transition electric dipole moment, and $\alpha_R(\om)$ is the real part of the polarizability at frequency $\om$ of a ground-state atom. Thus, for a single atom ($NV=1$),
\be
U= -\frac{2}{3\pi c^3}\sum_m\om_{mg}|\bd_{mg}|^2{\rm P}\int_0^{\Om}\frac{d\om\om^3}{\om_{mg}^2-\om^2},
\label{equ2}
\ee
where P denotes the principal value of the integral. In the case of a free, unbound electron with a nearly continuous energy spectrum, we can ignore $\om_{mg}^2$ compared with $\om^2$ in the denominator:
\be
U_{\rm free}=\frac{2}{3\pi c^3}\sum_m\om_{mg}|\bd_{mg}|^2\int_0^{\Om}d\om\om=\frac{e^2\hbar}{\pi mc^3}\int_0^{\Om}
d\om\om,
\ee
where we have used the Thomas--Reiche--Kuhn sum rule for the term multiplying the integral in the first equality. $U_{\rm free}$ is just the vacuum expectation value of $(e^2/2mc^2)\bA^2$, where $\bA$ is the (gauge-invariant) transverse vector potential. It can be subtracted away simply because it contributes the same amount to every energy level and therefore plays no role in measured transition frequencies in Lamb-shift experiments. Defining the observable shift of the ground state $g$ as $\Delta E_g=U-U_{\rm free}$, we obtain, for $\Om\gg \om_{mg}$ for all $\om_{mg}$,
\be
\Delta E_g= -\frac{2}{3\pi c^3}\sum_m\om_{mg}^3|\bd_{mg}|^2{\rm P}\int_0^{\Om}\frac{d\om\om}{\om_{mg}^2-\om^2} =\frac{2}{3\pi c^3}\sum_m\om_{mg}^3|\bd_{mg}|^2\log\frac{\Om}{\om_{mg}}.
\label{fp1}
\ee
Setting $\Om=mc^2/\hbar$, we have exactly the Bethe log formula for the Lamb shift. 

This derivation leads to both ``resonant" ($1/(\om_{mg}-\om)$ or ``on-shell" and ``non-resonant" ($1/(\om_{mg}+\om)$ or ``off-shell") energy denominators, corresponding respectively to real and virtual photon processes. Specifically, for $\Om\gg \om_{mg}$ for all $\om_{mg}$, the energy (\ref{fp1}) follows from
\be
-{\rm P}\int_0^{\Omega}\frac{d\om\om}{\om_{mg}^2-\om^2} =\frac{1}{2}\int_0^{\Om}\frac{d\om}{\om+\om_{mg}}+\frac{1}{2}{\rm P}\int_0^{\Om}\frac{d\om}{\om-\om_{mg}}\cong\frac{1}{2}\log\frac{\Om}{\om_{mg}}+\frac{1}{2}\log\frac{\Om}{\om_{mg}}.
\label{fp2}
\ee
{The presence of an on-shell contribution is somewhat disturbing since there are no real photon emission or absorption processes associated with resonant denominators for a ground-state atom in a vacuum of no photons.} In second-order perturbation theory the Lamb shift (before mass renormalization or relativistic corrections) is obtained from the one-loop process involving virtual (non-energy-conserving) transitions from the state in which an atom is in its ground state of energy $E_g$ and the field in its vacuum state, to intermediate states in which the atom is excited to a state of energy $E_m$ and there is one photon of energy $\hbar\om$ in the field. Only the non-resonant energy denominator $E_m-E_g+\hbar\om$ ($E_m>E_g$) appears in the integration over $\om$. (This contrasts with the energy shift for a state in which there is, for instance, a single photon in the field. Then there appears a resonant  denominator $E_m-E_g-\hbar\om$ because of the possible emission or absorption of the (real) photon.) The resonant denominator appearing in (\ref{eqexp}) is therefore extraneous.

 The resonant denominator is effectively eliminated by the high-frequency cutoff in the calculation above. Then, from Eq. (\ref{fp2}), the distinction between resonant and non-resonant terms disappears. Both then contribute, equally, to the Lamb shift. In the derivation of the Bethe log in Section \ref{sec:log}, in contrast, there is a complete, {\sl cutoff-independent} cancellation of the resonant denominator, leading to Bethe's formula.

 {For what follows it may be worth noting another interpretation of $U$ as defined by (\ref{equ2}). Consider
 \begin{align*}
 U_0=-\frac{1}{2}\big\langle\bd(t)\cdot\bE^{\perp}_0(t)\big\rangle,
 \end{align*}
 the energy of a dipole induced by a transverse electric field $\bE^{\perp}_0(t)$; the factor 1/2 accounts for the fact that the dipole is induced. Writing $\bd(t)$ as in (\ref{eqa1}), and proceeding as in Appendix B, we obtain the vacuum-field expectation value
 \begin{align*}
  U_0=-\frac{1}{2}\int_0^{\Omega}d\omega\alpha_R(\omega)
  \big\langle{\bE_{0\omega}^{\perp 2}}\big\rangle=U,   
 \end{align*}
 implying that $U$ can be regarded as a quadratic Stark shift induced by the vacuum field.}

 \section{Dipole Self-Interaction energy}\label{sec:self}
 {{The energy (\ref{eq1}) has been shown above to be the interaction $\big\langle -\frac{1}{2}\bd\cdot\bE^{\perp}_0\big\rangle$ of the atom with the source-free (transverse) electric field $\bE^{\perp}_0$; alternatively, it is the change in the zero-point electromagnetic energy caused by the presence of the atom. There is also an energy associated with the interaction of the atom with its own transverse source field $\bE_S^{\perp}$. Treating this interaction on the same footing as the interaction with the vacuum field, as an energy induced by a field acting on the atom, we define it as $U_S=\big\langle -\frac{1}{2}\bd\cdot\bE_S^{\perp}\big\rangle$.}

 The transverse electric dipole field $\bE_S^{\perp}(\br,t)$ of the atom is the complete electric field minus the unretarded, longitudinal (Coulomb) part. For an electric dipole moment (Hermitian operator) $\bd(t)=\hd d(t)$, the Heisenberg-picture operator
\be
\bE_S^{\perp}(\br,t)=-\frac{1}{c^2r}\sbkt{\hd-(\hd\cdot\hr)\hr}\ddot{d}\bkt{t-\frac{r}{c}}+\sbkt{\hd-3(\hd\cdot\hr)\hr}\sbkt{\frac{1}{cr^2}\dot{d}\bkt{t-\frac{r}{c}}+\frac{1}{r^3}d\bkt{t-\frac{r}{c}}}+\frac{1}{r^3}\sbkt{\hd-(\hd\cdot\hr)\hr}d(t),
\ee
which can be expressed as \cite{pwmvac}
\be
\bE_S^{\perp}(\br,t)=-\frac{1}{2\pi^2}\int d^3k[\hd-(\hd\cdot\hk)\hk]e^{i\bk\cdot\br}\int_0^tdt\pr\dot{d}(t\pr)\cos{\om_k(t\pr-t)},
\ee
where the circumflex denotes a unit vector. Then, for an atom at $\br=0$,
{{
\bea
U_S&=&-\frac{1}{2}\times\frac{1}{2}\big\la\bd(t)\cdot\bE_S^{\perp}(0,t)+\bE_S^{\perp}(0,t)\cdot\bd(t)\big\ra\nonumber\\
&=&\frac{1}{8\pi^2}\int d^3k[1-(\hd\cdot\hk)^2]\int_0^tdt\pr\big\la d(t)\dot{d}(t\pr)+\dot{d}(t\pr)d(t)\big\ra\cos\om_k(t\pr-t)\nonumber\\
&=&\frac{1}{3\pi c^3}\int_0^{\infty}d\om\om^2\int_0^tdt\pr\big\la d(t)\dot{d}(t\pr)+\dot{d}(t\pr)d(t)\big\ra\cos\om(t\pr-t).
\label{equ1}
\eea}
The dipole correlation fuction $\big\la d(t)\dot{d}(t\pr)+\dot{d}(t\pr)d(t)\big\ra$ follows from the zero-temperature fluctuation-dissipation theorem \cite{landau}:
\be
{\frac{1}{2}}\big\la d(t)\dot{d}(t\pr)+\dot{d}(t\pr)d(t)\big\ra = 
-\frac{\hbar}{\pi}\int_0^{\infty}dyy\alpha_I(y)\sin{y(t\pr-t)},
\label{eqv1}
\ee
where $\alpha_I(y)$ is the imaginary part of the atom's polarizability. A simplified derivation of this correlation function in the present context is given in Appendix B. Thus, from (\ref{equ1}),
{{
\be
U_S=-\frac{2\hbar}{\pi^2c^3}\int_0^{\infty}d\om\om^2\int_0^{\infty}dyy\alpha_I(y)\int_0^tdt\pr\sin{y(t\pr-t)}\cos{\om(t\pr-t)}
\rightarrow\frac{2\hbar}{\pi^2c^3}\int_0^{\Om}d\om\om^2{\rm P}\int_0^{\infty}\frac{dyy^2\alpha_I(y)}{y^2-\om^2},
\label{equ3}
\ee}
where we have introduced the high-frequency cutoff $\Om$ and have dropped terms arising from the artificial turning on at $t=0$ of the atom-field interaction. We have also multiplied by 3 to take into account the fact that $\big\la\bd(t)\cdot\dot{\bd}(t\pr)+\dot{\bd}(t\pr)\cdot\bd(t)\big\ra=3\big\la d(t)\dot{d}(t\pr)+\dot{d}(t\pr)d(t)\big\ra$ for the (spherically symmetric) atom.

\section{The Bethe log}\label{sec:log}
The change in the zero-point electromagnetic energy due to the presence of the atom is $U+U_S$, where $U$ is given by equation (\ref{equ2}) and $U_S$ by equation (\ref{equ3}):
\be
U+U_S=-\frac{\hbar}{\pi c^3}\int_0^{\Om}d\om\om^3\alpha_R(\om)+\frac{2\hbar}{\pi^2c^3}\int_0^{\Om}d\om\om^2{\rm P}\int_0^{\infty}\frac{dyy^2\alpha_I(y)}{y^2-\om^2},
\ee
where we have expressed $U$ in terms of the real part $\alpha_R(\om)$ of the atom's polarizability.
The fact that the complex polarizability $\alpha(\om)$ must be analytic in the upper half of the complex frequency plane allows us to express $\alpha_R(\om)$ in terms of $\alpha_I(\om)$ \cite{landau}:
\be
\alpha_R(\om)=\frac{2}{\pi}{\rm P}\int_0^{\infty}\frac{dyy\alpha_I(y)}{y^2-\om^2}.
\ee
Therefore,
\bea
U+U_S&=&-\frac{2\hbar}{\pi^2c^3}\int_0^{\Om}d\om\om^3{\rm P}\int_0^{\infty}\frac{dyy\alpha_I(y)}{y^2-\om^2}+\frac{2\hbar}{\pi^2c^3}\int_0^{\Om}d\om\om^2{\rm P}\int_0^{\infty}\frac{dyy^2\alpha_I(y}{y^2-\om^2}\nonumber\\
&=&\frac{2\hbar}{\pi^2c^3}\int_0^{\Om}d\om\om^2\int_0^{\infty}\frac{dyy\alpha_I(y)}{y+\om}.
\eea

The polarizabiliy $\alpha(\om)$ can be assumed to have the general form \cite{andrews,pwmloud}
\be
\alpha(\om)=\frac{1}{3\hbar}\sum_m|\bd_{mg}|^2\bkt{\frac{1}{\om_{mg}-\om-i\gamma_{mg}(\om)}+
\frac{1}{\om_{mg}+\om\pm i\gamma_{mg}(\om)}},
\ee
\be
\alpha_I(\om)=\frac{1}{3\hbar}\sum_m|\bd_{mg}|^2\bkt{\frac{\gamma_{mg}(\om)}{(\om_{mg}-\om)^2+\gamma^2_{mg}(\om)}\mp
\frac{\gamma_{mg}(\om)}{(\om_{mg}+\om)^2+\gamma^2_{mg}(\om)}},
\label{equ6}
\ee
with damping rates $\gamma_{mg}$, which can be assumed to be very small compared to the transition frequencies $\om_{mg}$. Therefore the dominant contribution to $U+U_S$ comes from the first term in parentheses in (\ref{equ6}), and, since this term is strongly peaked at $y=\om_{mg}$, {{we make the approximation of replacing $\gamma_{mg}(\om)$ by $\gamma_{mg}(\om_{mg})$ and extending the $y$ integration to $-\infty$}}:
\be
U+U_S\cong\frac{2}{3\pi^2c^3}\sum_m|\bd_{mg}|^2\int_0^{\Om}d\om\om^2\frac{\om_{mg}}{\om_{mg}+\om}\int_{-\infty}^{\infty}dy
\bkt{\frac{\gamma_{mg}(\om_{mg})}{(\om_{mg}-y)^2+\gamma_{mg}^2(\om_{mg})}}
= \frac{2}{3\pi c^3}\sum_m\om_{mg}|\bd_{mg}|^2\int_0^{\Om}\frac{d\om\om^2}{\om_{mg}+\om},
\label{equ7}
\ee
where we have again introduced the high-frequency cutoff $\Om$ in this nonrelativistic calculation. It may be worth noting that this result has no contribution from the last term in (\ref{equ6}). In other words, it has not been necessary to address the question of the sign of the damping term in the non-resonant part of the polarizability \cite{andrews,pwmloud}.

As in the preceding section we now subtract the free-electron energy $U_{\rm free}$:
\be
U+U_S-U_{\rm free}=-\frac{2}{3\pi c^3}\sum_m\om^2_{mg}|\bd_{mg}|^2\int_0^{\Om}\frac{d\om\om}{\om_{mg}+\om}.
\label{equ8}
\ee
Finally we derive the ``observable" shift in the energy level $E_g$ by Bethe's mass renormalization. For our purposes this is tantamount to subtracting from (\ref{equ8}) the energy obtained from (\ref{equ8}) by putting $\om_{mg}=0$ in the integrand. The result,
\be
\Delta E_g=\frac{2}{3\pi c^3}\sum_m\om_{mg}^3|\bd_{mg}|^2\int_0^{\Om}\frac{d\om}{\om_{mg}+\om}=
\frac{2}{3\pi c^3}\sum_m\om_{mg}^3|\bd_{mg}|^2\log\frac{\Om}{\om_{mg}},
\ee
with $\Om=mc^2/\hbar$, is the Bethe log.

\section{Discussion}\label{sec:disc}
A physical interpretation of Bethe's formula as a consequence of the quantum fluctuations of the electromagnetic field was discussed in the influential and still frequently cited paper of Welton \cite{welton}. We have followed the interpretation suggested by Feynman in which the Lamb shift is regarded as a change in electromagnetic energy due to the presence of the atom. This interpretation, though not nearly as influential as Welton's, is more in line with the usual explanation of the Casimir force between conducting plates, for example, in terms of the change in field energy caused by the presence of the plates. 

As discussed in Section \ref{sec:feyn}, however, a calculation based on Feynman's original suggestion involves a contribution from a real, ``on-shell" photon process. {{Such a contribution appears because of the refractive index [equation (\ref{eqf})] and, as in Power's analysis \cite{power}, its expression in terms of the Kramers--Heisenberg formula. But the Kramers--Heisenberg formula is derived for the case of scattering of real, incident photons \cite{refs}, and therefore includes an on-shell contribution which, in the case of the Lamb shift,  would imply the possibility of absorption from the vacuum of no photons.}}
This spurious contribution is effectively eliminated, as shown in Section \ref{sec:feyn}, by the high-energy cutoff in the nonrelativistic calculation. We have modified Feynman's argument by re-defining the electromagnetic energy in that argument to include  the self-interaction of the atom with itself. The Lamb shift is then regarded as the difference between this total energy and the field energy in free space. Then the unphysical on-shell contribution is exactly cancelled, independently of the cutoff. 

We have restricted the treatment here to ground-state atoms. This avoids consideration of excited-state spontaneous emission, which would only complicate our analysis somewhat without affecting our main points. 

Finally we note that Passante and Rizzuto \cite{pass} have shown how effective Hamiltonians can be formulated to exclude on-shell contributions of the type considered in this paper.

{
{\section*{Acknowledgment}
PWM thanks G. Jordan Maclay for discussions relating to the Lamb shift and the Bethe log formula. The research of PRB is supported by the Air Force Office of Scientific Research and the National Science Foundation. KS acknowledges support from the National Science Foundation under Grant No. PHY-2309341, and by the John Templeton Foundation under Award No. 62422. This research was supported in part by grant NSF PHY-2309135 to the Kavli Institute for Theoretical Physics (KITP).}}
\appendix
\section{Another derivation of Equation (\ref{eq1})}
Equation (\ref{eq1}) follows from the expectation value of the total electromagnetic energy density in the absence of any dissipation: 
\be
u_{\om}=\frac{1}{8\pi}\sbkt{\frac{d}{d\om}(\eps\om)\big\la\bE_{\om}^2\big\ra+\big\la\bH_{\om}^2
\big\ra}=\frac{1}{8\pi}\sbkt{\frac{d}{d\om}(\eps\om)\big\la\bE_{\om}^2\big\ra+\eps(\om)\big\la\bE_{\om}^2\big\ra}=\frac{n}{4\pi}\frac{d}{d\om}(n\om)\big\la\bE_{\om}^2\big\ra,
\label{eqaa1}
\ee
where $\eps(\om)=n^2(\om)$. From the formula \cite{diel}
\be
\bE(\br,t)=i\sum\sss\bkt{\frac{2\pi\hbar\om_k}{Vn(dn/d\om)}}^{1/2}\sbkt{a\sss e^{i\bk\cdot\br}-\ad\sss e^{-i\bk\cdot\br}}\bek
\label{efield}
\ee
for the quantized electric field in such a medium we obtain the vacuum-field expectation value
\be
\big\la\bE^2(\br,t)\big\ra=\sum\sss\frac{2\pi\hbar\om_k}{Vn(dn/d\om)}\big\la a\sss \ad\sss\big\ra=\frac{2\hbar}{\pi c^3}\int_0^{\Om}d\om\om^3n(\om)
\ee
in the mode-continuum limit ($\sum\sss\rightarrow (2V/8\pi^3)\int d^3k$), implying
\be
\big\la\bE_{\om}^2\big\ra=\frac{2\hbar}{\pi c^3}\om^3n(\om).
\ee
Then from (\ref{eqaa1}) we obtain, after integration over all frequencies and multiplication by the volume $V$,
\be
{\blue V\int_0^{\infty}d\om u_{\omega}}=\frac{\hbar V}{2\pi^2 c^3}\int_0^{\Om}d\om\om^3 n^2(\om)\frac{d}{d\om}(n\om),
\ee
in agreement with the energy (\ref{eq1}).

\section{Dipole correlation function}
The electric dipole moment induced by the quantized, ``external" electric field at the position $\br=0$ of the atom is expressed in linear response theory as
\be
\bd(t)=i\sum\sss\bkt{\frac{2\pi\hbar\om_k}{V}}^{1/2}
\sbkt{\alpha(\om_k)a\sss e^{-i\om_kt}-\alpha^*(\om_k)\ad\sss e^{i\om_kt}}\bek,
\label{eqa1}
\ee
where $a\sss$ and $\ad\sss$ are the usual photon annihilation and creation operators 
and $\bek$  (assumed real) is a unit polarization vector for the plane-wave mode with wave vector $\bk$ and polarization index $\lambda$. In the case of interest here the electric field is the source-free field with vacuum expectation values $\la a\sss a_{\bk\pr\lambda\pr}\ra=\la\ad\sss a_{\bk\pr\lambda\pr}\ra=\la \ad\sss \ad_{\bk\pr\lambda\pr}\ra=0$ and $\la a\sss \ad_{\bk\pr\lambda\pr}\ra
=\delta_{\bk,\bk\pr}\delta_{\lambda,\lambda\pr}$. It then follows from (\ref{eqa1}) that
\be
\frac{1}{2}\big\la\bd(t)\cdot{\bd}(t\pr)+{\bd}(t\pr)\cdot\bd(t)\big\ra=\frac{2\hbar}{\pi c^3}\int_0^{\infty}d\om\om^3|\alpha(\om)|^2\cos{\om(t\pr-t)}
\ee
in the mode-continuum limit. From the optical theorem for Rayleigh scattering or radiative damping \cite{pwmloud},
\be
\alpha_I(\om)=\frac{2\om^3}{3c^3}|\alpha(\om)|^2,
\ee
and therefore
\be
\frac{1}{2}\big\la\bd(t)\cdot{\bd}(t\pr)+{\bd}(t\pr)\cdot\bd(t)\big\ra=3\times\frac{1}{2}
\big\la d(t)\cdot{d}(t\pr)+{d}(t\pr)\cdot d(t)\big\ra
=\frac{3\hbar}{\pi}\int_0^{\infty}d\om\alpha_I(\om)\cos{\om(t\pr-t)}.
\ee
Equation (\ref{eqv1}) follows by differentiation.

\end{document}